\documentclass[final,5p,times,twocolumn]{elsarticle} 

\usepackage{lineno,hyperref}
\usepackage{hyperref}
\usepackage{siunitx}
\modulolinenumbers[5]

\journal{Journal of \LaTeX\ Templates}

\begin{document}

\begin{frontmatter}

\title{Results from the EPICAL-2 Ultra-High Granularity  Electromagnetic Calorimeter Prototype}

\author[a]{T. Peitzmann\corref{mycorrespondingauthor}}
\cortext[mycorrespondingauthor]{email: t.peitzmann@uu.nl}
\author[b]{J. Alme}
\author[a]{R. Barthel}
\author[a]{A. van Bochove}
\author[c,d]{V. Borshchov}
\author[e]{R. Bosley}
\author[a]{A. van den Brink}
\author[a]{E. Broeils}
\author[f]{H. B\"usching}
\author[b]{V.N. Eikeland}
\author[b]{O.S. Groettvik}
\author[g]{Y.H. Han}
\author[a,h]{N. van der Kolk}
\author[g]{J.H. Kim}
\author[g]{T.J. Kim}
\author[g]{Y. Kwon}
\author[i]{M. Mager}
\author[j]{Q. W. Malik}
\author[a]{E. Okkinga}
\author[g]{T.Y. Park}
\author[f]{F. Pliquett}
\author[c,d]{M. Protsenko}
\author[i]{F. Reidt}
\author[a]{S. van Rijk}
\author[j]{K. R{\o}ed}
\author[f]{T.S. Rogoschinski}
\author[b]{D. R\"ohrich}
\author[a]{M. Rossewij}
\author[a]{G.B. Ruis}
\author[b,j]{E. H. Solheim}
\author[c,d]{I. Tymchuk}
\author[b]{K. Ullaland}
\author[e]{N. Watson}
\author[a,h]{H. Yokoyama}



\address[a]{Institute for Gravitational and Subatomic Physics (GRASP), Utrecht University/Nikhef, Utrecht, Netherlands}
\address[b]{Department of Physics and Technology, University of Bergen, Bergen, Norway} 
\address[c]{Research and Production Enterprise ``LTU'' (RPE LTU), Kharkiv, Ukraine}
\address[d]{Bogolyubov Institute for Theoretical Physics, Kyiv, Ukraine}
\address[e]{School of Physics and Astronomy, University of Birmingham, Birmingham, United Kingdom}
\address[f]{Institut f\"ur Kernphysik, Johann Wolfgang Goethe-Universit\"at Frankfurt, Frankfurt, Germany}
\address[g]{Yonsei University, Seoul, Republic of Korea}
\address[h]{Nikhef, National Institute for Subatomic Physics, Amsterdam, Netherlands}
\address[i]{European Organization for Nuclear Research (CERN), Geneva, Switzerland}
\address[j]{Department of Physics, University of Oslo, Oslo, Norway}


\begin{abstract}
A prototype of a new type of calorimeter has been designed and constructed, based on a silicon-tungsten sampling design using pixel sensors with digital readout. It makes use of the Alpide  sensor developed for the ALICE Inner Tracking System (ITS) upgrade. A binary readout is possible due to the pixel size of $\approx 30 \times 30 \, \mu \mathrm{m}^2$. This prototype has been successfully tested with cosmic muons and with test beams at DESY and the CERN SPS.
We report on performance results obtained at DESY, showing good energy resolution and linearity, and compare to detailed MC simulations. Also shown are preliminary results of the high-energy performance as measured at the SPS. The two-shower separation capabilities are discussed.
\end{abstract}


\end{frontmatter}


\section{Introduction}
The paradigm of digital calorimetry is the measurement of a high-energy particle by counting shower particles. 
Proof-of-principle studies have been performed with the  \textsc{Epical-1} prototype based on MIMOSA pixel sensors and are reported in \cite{deHaas:2017fkf}. In these proceedings, we report results from the next-generation prototype, \textsc{Epical-2}, which uses ALPIDE sensors. The project is closely related to an ongoing development of a protonCT detector \cite{Alme2020} and uses the same technology for the active layers containing the ALPIDE sensors.
It simultaneously serves as fundamental R\&D on digital calorimetry and as a contribution to provide one of the technologies for the FoCal detector proposed \cite{Focal-loi} for the ALICE experiment at the CERN LHC.

\begin{figure}[hbt]
     \begin{center}
      \includegraphics[width=0.9\columnwidth]{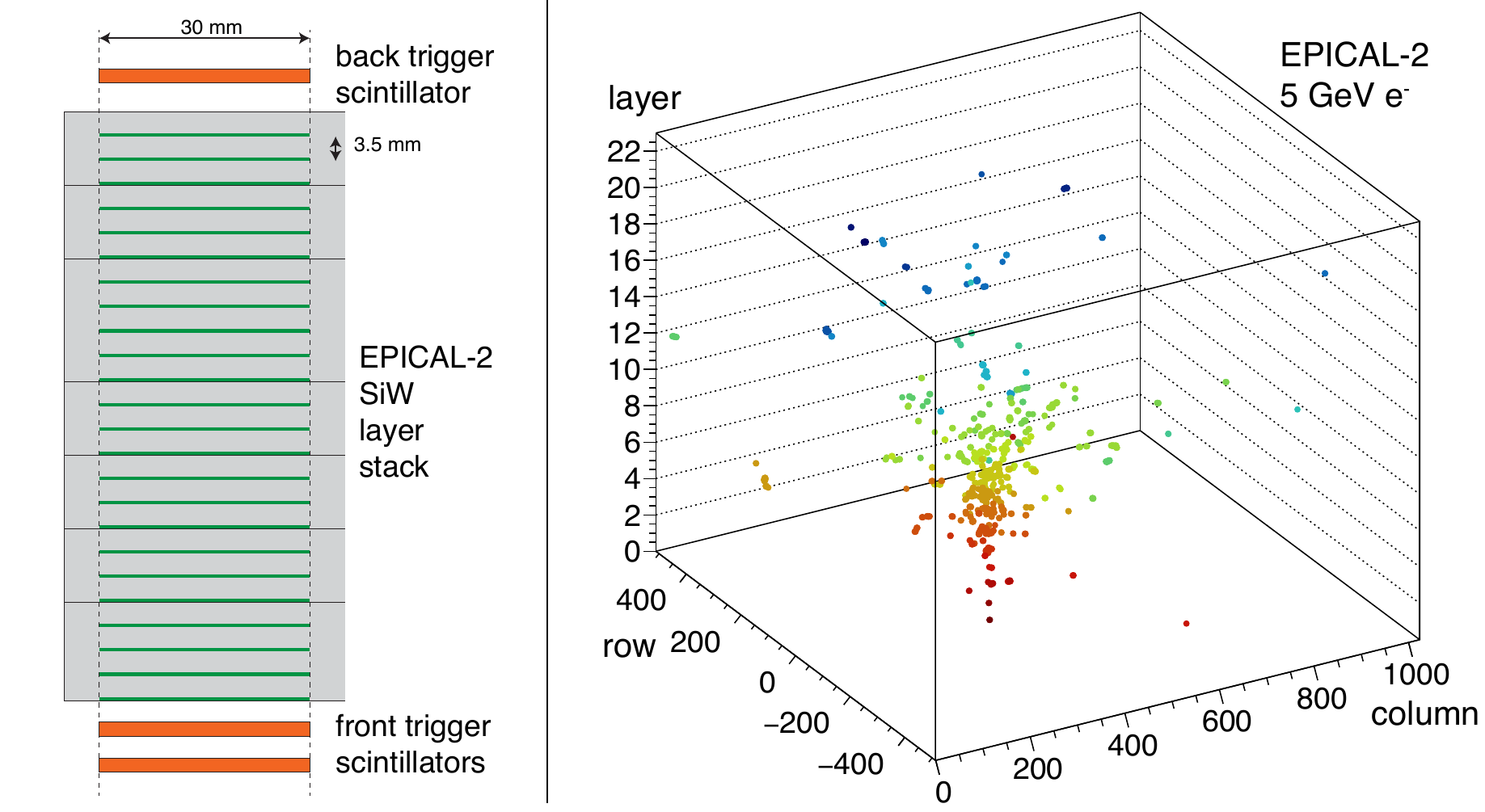}
    \end{center}
  \caption{Left: Schematic view of the \textsc{Epical-2} detector stack showing the W absorber (grey), the sensitive layers (green), and the trigger scintillators (orange). Right: Measured spatial distribution of hits in \textsc{Epical-2} for a single 5 GeV electron event. The colours of the dots correspond to the layer number, with red corresponding to front layers and blue to back layers.}
  \label{fig:event}
\end{figure}

The new prototype is a SiW sampling detector with 24 layers of $3 \times 3 \, \mathrm{cm}^2$ active cross section and a thickness of  approximately $ 0.86 X_0$ each, consisting of tungsten absorbers and pairs of Si sensors. It utilises the ALPIDE CMOS pixel sensors developed at CERN for the ALICE ITS2 \cite{ALPIDE}, which consists of a  matrix of $1024 \times 512$ pixels of area \SI[product-units=power]{29.24 x 26.88}{\micro\m}. The state-of-the-art sensors allow for a readout speed of \textsc{Epical-2} suitable for a modern particle physics experiment like ALICE, unlike the previous prototype. As the sensor is designed for tracking, one of the aims of this project is to study its suitability for calorimetry. The individual sensors of the detector have been calibrated and aligned using measurements with cosmic muons. Here, a selection of results on the test beam performance of \textsc{Epical-2} is presented, from measurements at DESY in 2019, and from a preliminary analysis of data taken at the SPS in 2021. For comparison, detailed MC simulations have been performed with ${\mathrm{Allpix^2}}$\cite{SPANNAGEL2018164}.
 More details on the detector and the analysis can be found in an upcoming publication \cite{epical2}.

Fig.~\ref{fig:event} shows an event display of a measurement of a single electron of 5~GeV. The fired pixels are shown at their position as a function of depth in terms of layer number and of transverse dimensions in terms of pixel row and column. One can clearly observe the characteristic development of a shower, with a narrow distribution in early layers, a maximum density after a few layers and a broadening in later layers. The average behaviour of the spatial distributions is consistent with the expectations from simulations.
 It is also apparent that fluctuations are significant for this low energy. 
The exploitation of the rich information content of these measurements has only started.
 
\begin{figure}[htb]
     \begin{center}
      \includegraphics[width=0.85\columnwidth]{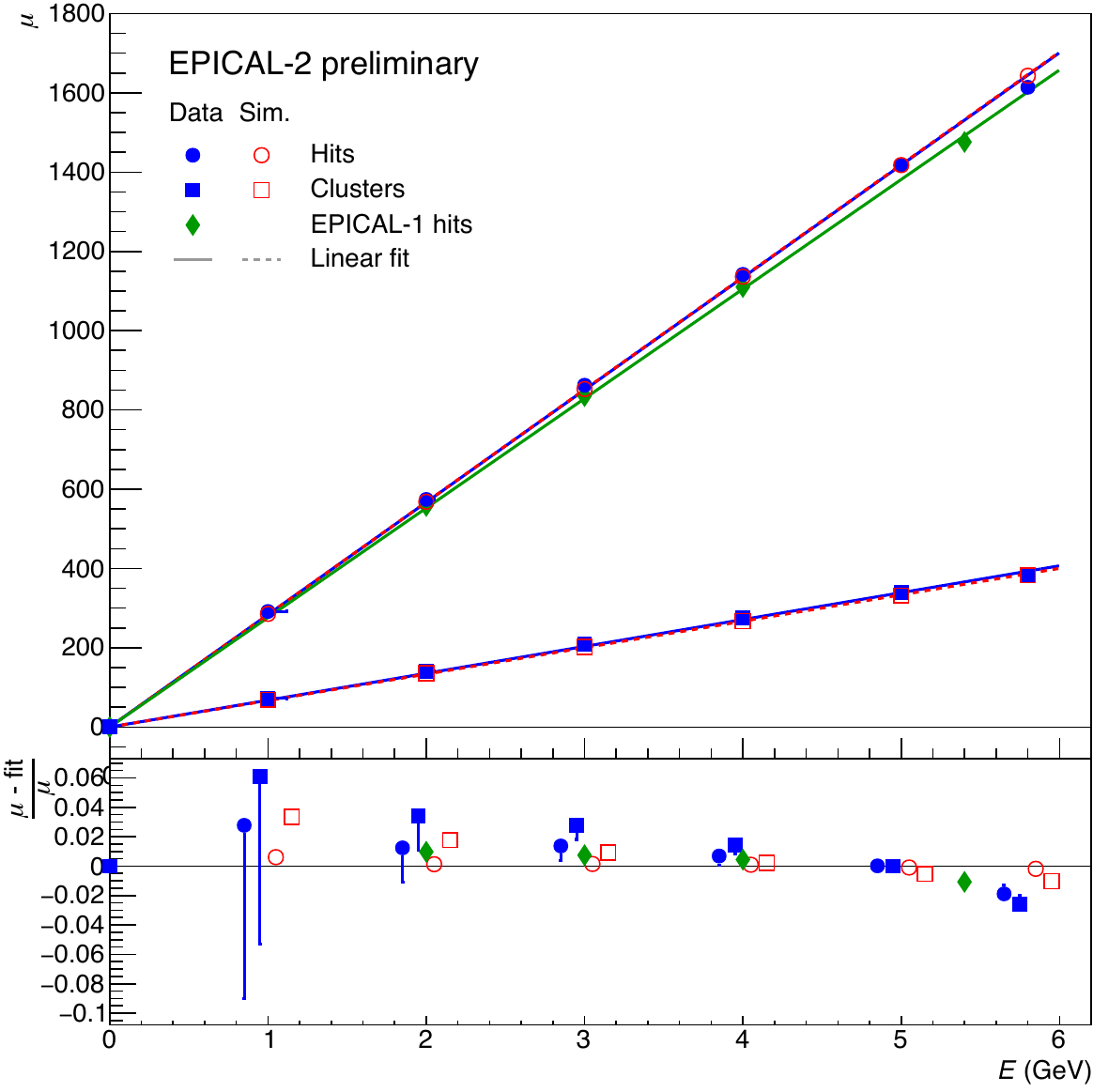}
    \end{center}
  \caption{Top: Energy response in terms of $N_{\mathrm{hits}}$ (circles) and  $N_{\mathrm{clus}}$ (squares) as a function of true energy from the DESY test beam, for experimental data (filled blue symbols) and simulation (open red symbols). 
 Data on $N_{\mathrm{hits}}$ from the \textsc{Epical-1} prototype are shown for comparison as green markers. 
  Bottom: Relative difference of data points to linear fits. In these ratios for \textsc{Epical-2}, a one-sided systematic error related to the beam energy uncertainty is included.}
  \label{fig:linearity}
\end{figure}
\begin{figure}[hbt]
      \begin{center}
      \includegraphics[width=0.575\columnwidth]{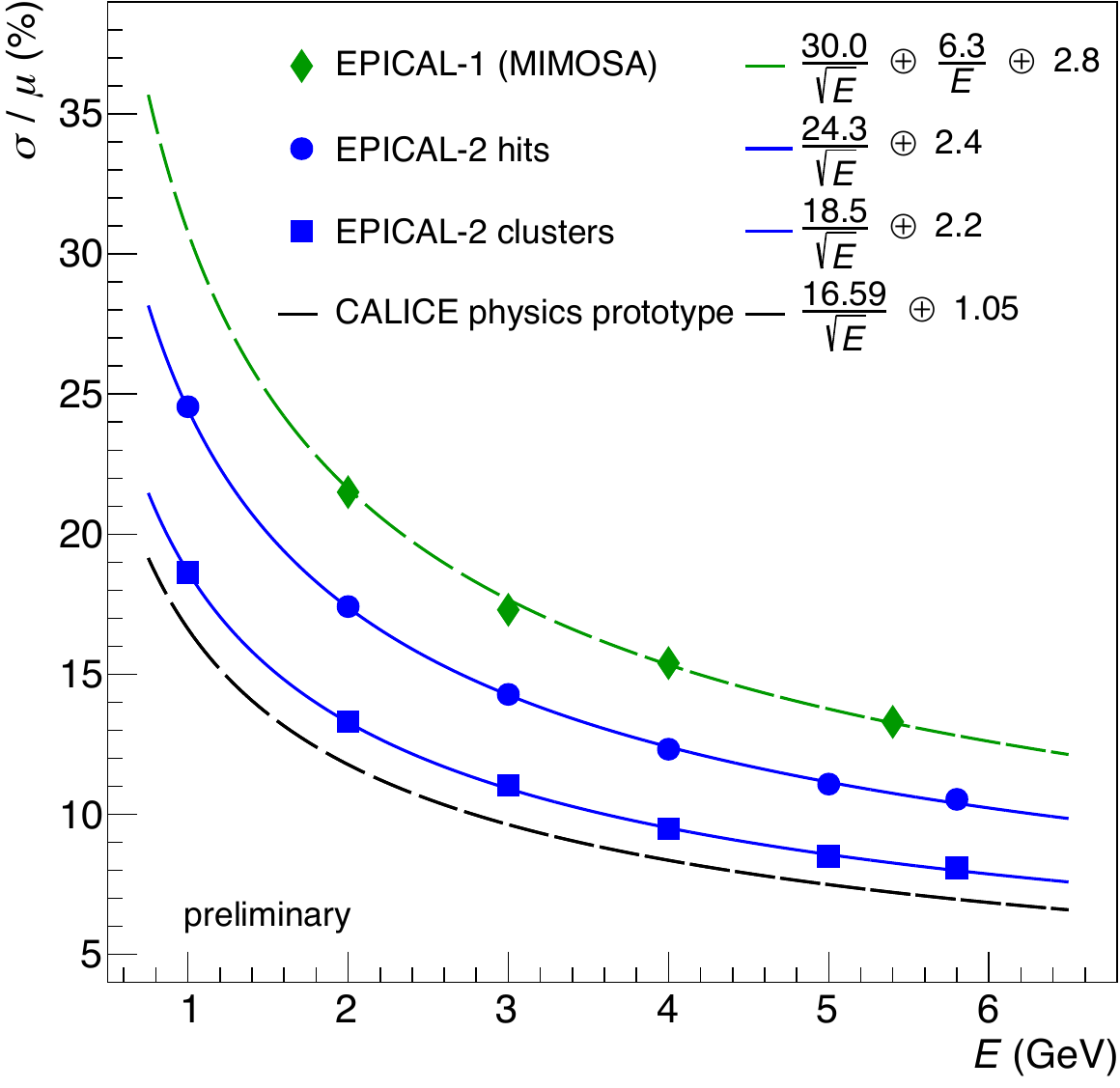}
    \end{center}
  \caption{Energy resolution as a function of beam energy. The results for experimental data are shown as filled blue circles for $N_{\mathrm{hits}}$ and as filled blue squares for $N_{\mathrm{clus}}$. 
 Included is a comparison to results from the \textsc{Epical-1} prototype \cite{deHaas:2017fkf} (green diamonds) and to the parameterisation of the resolution of the CALICE silicon-tungsten ECAL physics prototype \cite{CALICE:2008kht} (black dashed line).}
  \label{fig:resolution}
\end{figure}

\section{Performance at Low Energy}

Both the number of hits (fired pixels) $N_{\mathrm{hits}}$ and the number of clusters (groups of adjacent hits) $N_{\mathrm{clus}}$ have been used as observables for the energy response. Fig.~\ref{fig:linearity} shows the mean value of these observables as a function of beam energy. Naturally, $N_{\mathrm{clus}}$ is lower than $N_{\mathrm{hits}}$. Both observables show a linear rise with beam energy, and the MC simulation agrees well with the experimental data. The data points are fitted with a linear function constrained to an intercept of 0, which for experimental data is taken from pedestal measurements. The lower panel of Fig.~\ref{fig:linearity} shows the relative difference of the data points to the corresponding fit. The results from simulation for $N_{\mathrm{hits}}$ agree perfectly with this linear fit. The measured data for the same observable show a small deviation from linearity.  As there is a possible deviation of the mean value of the true energy \cite{DESY,priv:stanitzki}, we have attempted to incorporate this  into error bars, but these are difficult to estimate in particular at the highest energy. This suggests that the observed non-linearity is to a large extent related to the test beam properties. This is further supported by the agreement of similar measurements with a different detector (green symbols for EPICAL-1). In both experimental data and simulations, $N_{\mathrm{clus}}$ shows a slightly larger non-linearity, which is very likely due to cluster overlap. 

The energy resolution is shown in Fig~\ref{fig:resolution} and is compared to two other experimental results. The resolution obtained from $N_{\mathrm{hits}}$ is already significantly better than the one from the previous prototype, as expected because that detector suffered from a number of non-working sensors. Interestingly, the resolution obtained from $N_{\mathrm{clus}}$ is significantly better than the one from $N_{\mathrm{hits}}$. Apparently, the observable $N_{\mathrm{clus}}$ is less sensitive to single particle response fluctuations, which lead to variations of the cluster size, and thus to additional fluctuations in $N_{\mathrm{hits}}$. The data are fitted with the function traditionally used to describe the energy dependence of the resolution of calorimeters:
\begin{equation}
\frac{\sigma_E}{E} = \frac{a}{\sqrt{E/\mathrm{GeV}}}\oplus b \oplus \frac{c}{E/\mathrm{GeV}},
\label{eq:resolution}
\end{equation}
and the results are indicated in the figure. The fits agree well with the data. The noise term $c$ is not needed for \mbox{\textsc{Epical-2}}, because, unlike for \textsc{Epical-1}, the noise is negligible. Finally, the data are also compared to results from the CALICE silicon-tungsten ECAL physics prototype \cite{CALICE:2008kht}, a state-of-the-art calorimeter with analog readout, which shows only slightly better resolution than our prototype. It should be noted that the present results are not corrected for the intrinsic momentum spread of the test beam.

\section{Performance at High Energy}

\begin{figure}[hbt]
     \begin{center}
      \includegraphics[width=0.9\columnwidth]{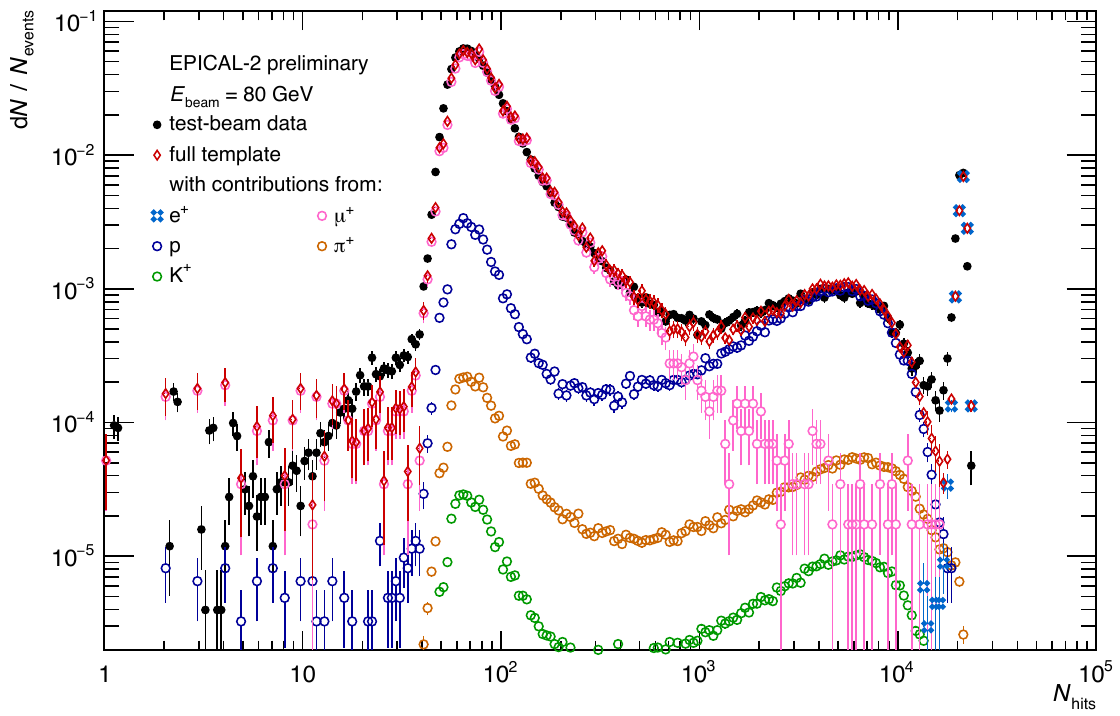}
    \end{center}
  \caption{Distribution of the number of hits for a mixed beam of 80 GeV at the CERN SPS H6 beam line. The black symbols show the measured data, symbols of different colours the contributions of different particle species from simulation. The red symbols show the sum of all simulation contributions.}
  \label{fig:beam-composition}
\end{figure}

Recently, measurements with the detector prototype have also been performed with a mixed particle beam at the CERN SPS. The $N_{\mathrm{clus}}$  observable shows already first signs of non-linearity at low energy, and the effect of cluster overlap is expected to become more important for increasing hit density, and thus implicitly for higher beam energy. Therefore, for a first analysis of high energy measurements, only $N_{\mathrm{hits}}$ has been used. Fig.~\ref{fig:beam-composition} shows the distributions of $N_{\mathrm{hits}}$ at a beam energy of 80~GeV. Three different maxima are observed: at $N_{\mathrm{hits}} \approx 70$ there is a minimum-ionising peak with a Landau-like tail to large values, there is a broad peak around $N_{\mathrm{hits}} \approx 6000$ due to showering hadrons and a sharp peak for $N_{\mathrm{hits}} \approx 20000$ due to electrons. This interpretation is supported by the decomposition of the spectrum into the contributions for different particles as predicted by the MC simulations and shown in the figure. The description by simulation is very good, even to the level of very rare events with particles producing only very few hits seen to the left of the MIP. It is apparent that the electron component is well separated, so that an independent analysis is possible. For the results below, we restrict the analysis to a $3 \sigma$ interval around the electron peak, where the hadron contamination should be of the order of $1\%$ or smaller. 

\begin{figure}[hbt]
     \begin{center}
      \includegraphics[width=0.8\columnwidth]{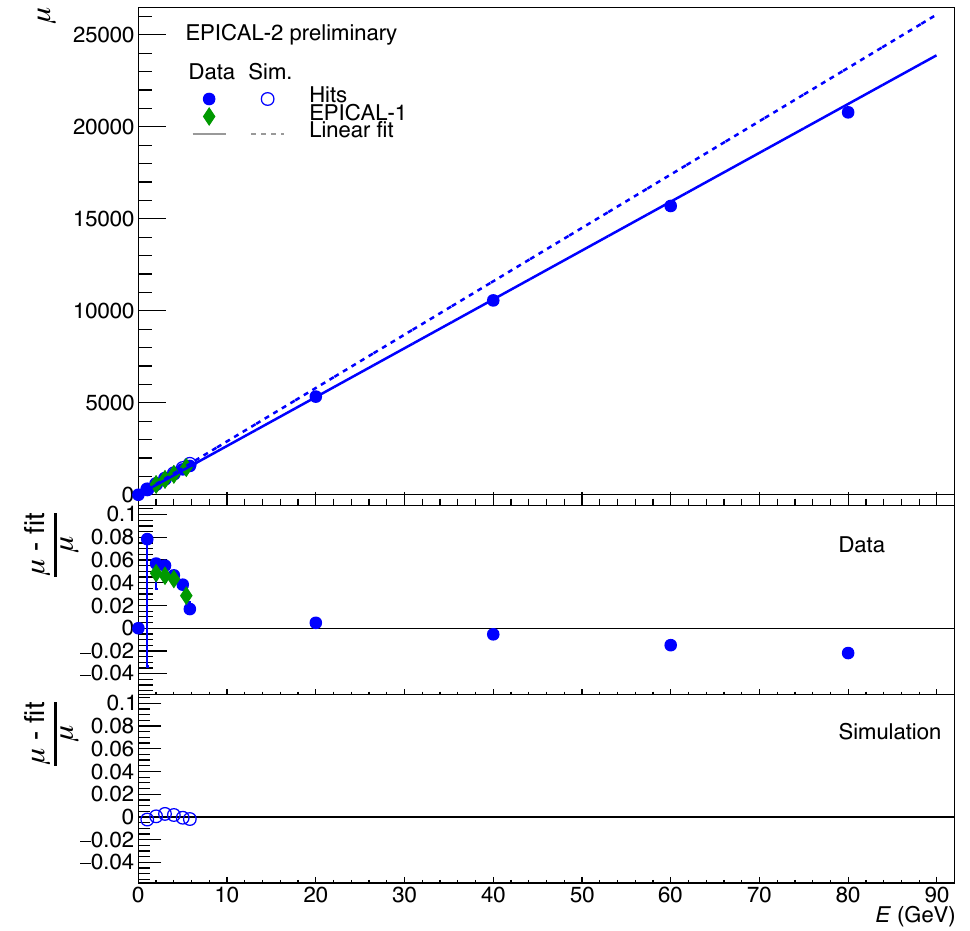}
    \end{center}
  \caption{Top: Energy response in terms of $N_{\mathrm{hits}}$ (squares) as a function of energy. Data from the \textsc{Epical-1} prototype are shown for comparison as green markers. Bottom: Relative difference of data points to the linear fits.}
  \label{fig:linearity2}
\end{figure}

\begin{figure}[hbt]
      \begin{center}
      \includegraphics[width=0.55\columnwidth]{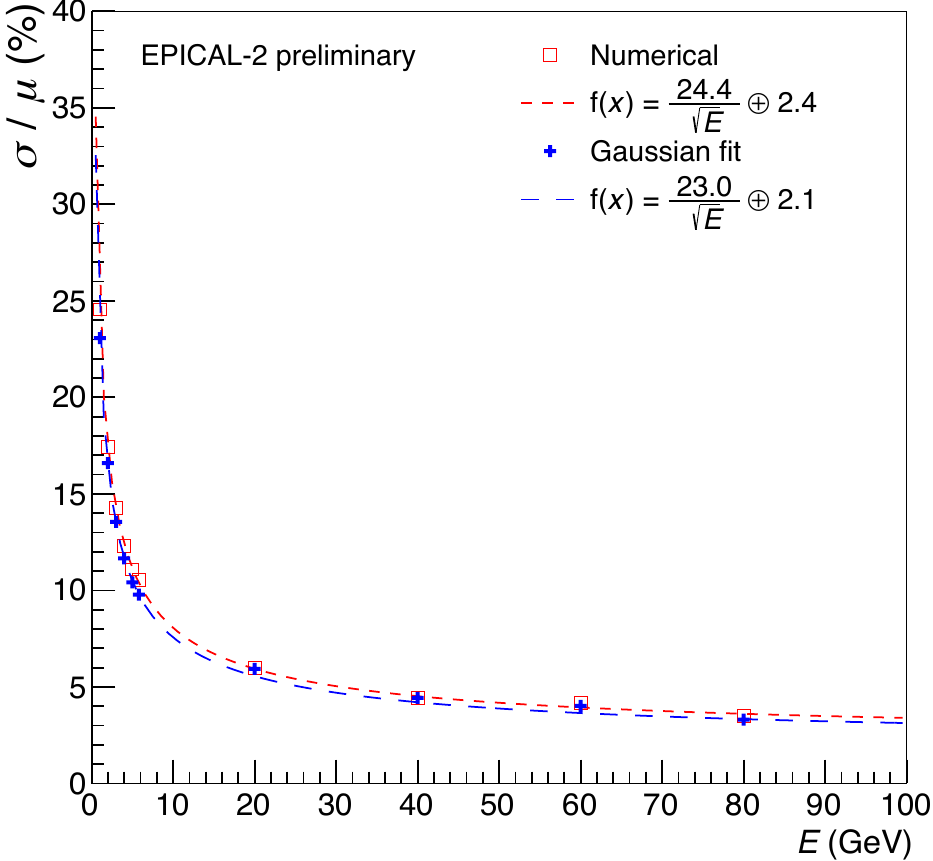}
    \end{center}
  \caption{Energy resolution as determined from $N_{\mathrm{hits}}$ as a function of beam energy.}
  \label{fig:resolution2}
\end{figure}

 Fig.~\ref{fig:linearity2} shows the mean value of the response as a function of beam energy. Now, in addition to the results discussed above, preliminary results from measurements at the SPS H6 test beam up to $E_{\mathrm{beam}} = 80 \, \mathrm{GeV}$ are shown. Again, linear fits are performed. As only low-energy data points were available from simulations at this point, the corresponding linear fit is the same as in Fig.~\ref{fig:resolution}, while the fit to measured data has changed due to the additional data points. The bottom panel again shows the relative deviation from linearity -- the stronger deviation of the measured low energy points is due to the different fit used here. 

Again, data show good linearity when taking into account that the non-linearity at low energy is mostly due to the beam energy uncertainty. The small deviations from linearity at high energy could be due to saturation of the apparent hit density in the shower core. We expect that this can be corrected on an event-by-event basis, but such a study is beyond the scope of these proceedings. Likely, a similar correction can be performed for $N_{\mathrm{clus}}$ to make that observable also usable at high energy.

Fig.~\ref{fig:resolution2} shows the energy resolution when using $N_{\mathrm{hits}}$, now also extended to higher energy with preliminary results. The behaviour of this observable from low energy is confirmed, demonstrating good energy resolution. This analysis is likely more sensitive to a possible contamination by hadrons, but we expect further corrections to be small.

\section{High-Granularity Potential}

\begin{figure}[hbt]
      \begin{center}
      \includegraphics[width=1\columnwidth]{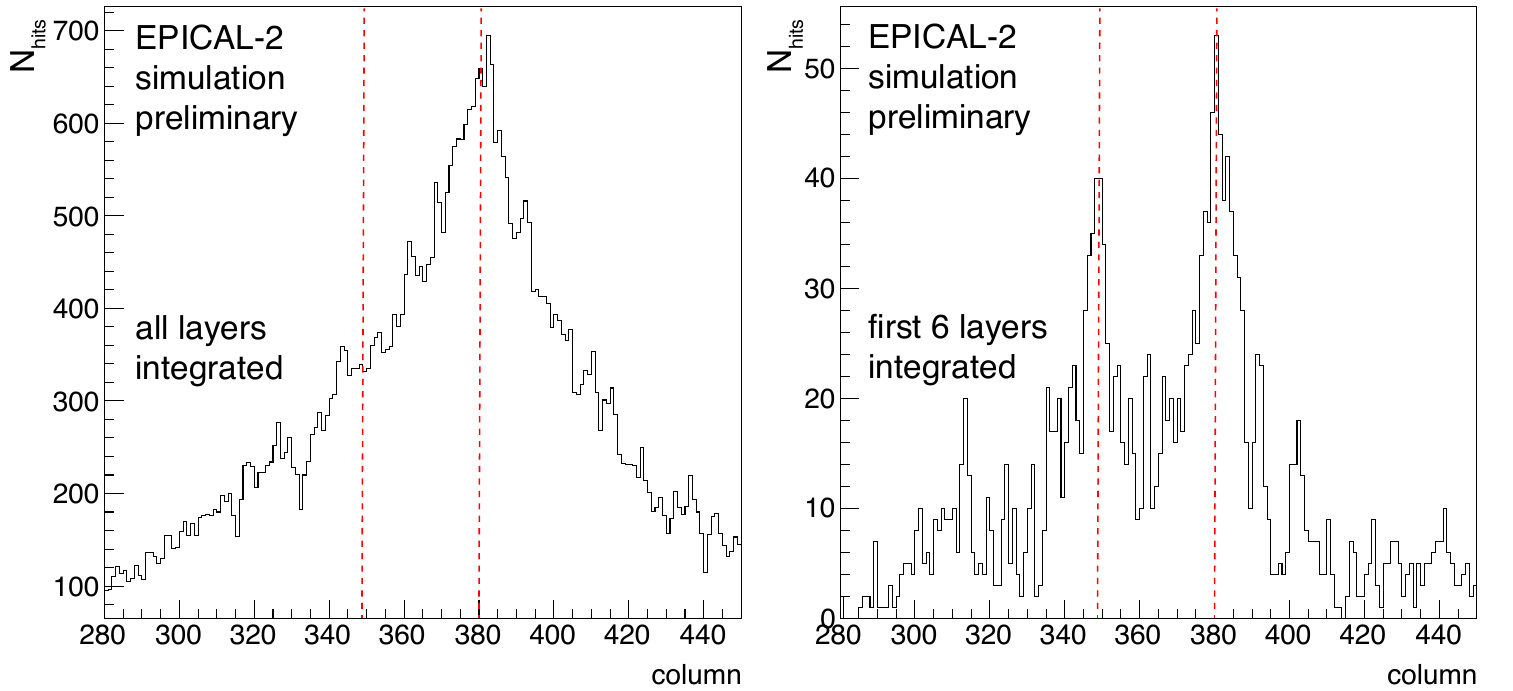}
    \end{center}
  \caption{Distributions of the number of hits for a single event from MC simulation containing two electrons with a separation of $\approx 1.2 \, \mathrm{mm}$. The distributions are integrated over one transverse dimension and over either all layers (left panel) or the first 6 layers (right panel). The dotted red lines indicate the position of incidence of the two particles.}
  \label{fig:twoshower}
\end{figure}

While the above results demonstrate that the general calorimetric performance of the new technology is good, they do not make full use of its particular strength. Detailed information of the event-by-event distributions of showers (as shown e.g. Fig.~\ref{fig:event}) is available, but analysis exploiting this is still ongoing. Usage of that information may allow further improvement of the reconstruction of showers. It will in particular allow the separation of overlapping showers at very small distances. This can be demonstrated from simulations of such a case, as shown in Fig.~\ref{fig:twoshower}, where hit distributions are given for a single simulated event containing one electron of 250~GeV and a second one of 30~GeV vertically separated by $\approx 1.2 \, \mathrm{mm}$. The distributions are shown as a function of pixel column integrated over the other transverse coordinate (pixel row). For the left panel, the distributions were integrated over the full depth of the detector, i.e. all sensor layers. Just a single peak of the more highly energetic electron is observed here, no hint of the lower energy electron is visible. For the right panel, only the first 6 layers were used for  longitudinal integration. In this case, two peaks are observed, one each at the position of incidence of one of the incoming particles. Clearly, already with such a simple measure, \mbox{\textsc{Epical-2}} can easily discriminate two close-by showers at distances of the order of 1~mm, which would be perceived as a single shower in a conventional calorimeter. Using individual layers, the digital pixel technology will ensure this kind of performance also for the future FoCal detector in ALICE.

A fully digital pixel calorimeter, like \textsc{Epical-2}, will allow one to perform much more powerful analysis of showers. An analysis of the three-dimensional shower distribution is possible event by event, likely pushing the two-shower separation power to even smaller distances, possibly improving the energy measurement beyond what is shown here via an explicit correction for saturation and allowing for the subtraction of e.g. the contribution of a minimum-ionizing particle overlapping an electromagnetic shower. More in general, the device should have extremely high potential for the use of particle flow algorithms \cite{Brient:2001fow}.

\section{Conclusion}
\textsc{Epical-2}, a prototype of an electromagnetic digital pixel calorimeter, has been constructed and successfully tested with cosmics and in test beams. The detector has performed extremely well, with low noise and good uniformity. The chosen sensor, the ALPIDE, has shown its suitability for the calorimetric environment. Simulations using the ${\mathrm{Allpix^2}}$ framework describe the behaviour of the device up to a high level of detail.

Results from a detailed analysis of low-energy measurements are presented. The basic calorimetric performance, in terms of energy linearity and resolution, is very good. Of the two observables employed, $N_{\mathrm{clus}}$ provides better resolution, but is slightly worse in terms of linearity. Efforts are under way to find the optimal energy reconstruction. With the current algorithms, the resolution at low energy is comparable to the state of the art of analog SiW calorimeters.

Results at high energy, as observed from a preliminary analysis, are consistent with the behaviour at low energy, confirming the good performance. These results are obtained using a mixed particle beam, so a hadron contamination is not fully excluded, although we expect it to be very small. Further advancements of reconstruction algorithms may be more important to achieve optimum performance for high-energy showers. The current knowledge gives confidence that the technology will meet the requirements for pixel layers in FoCal. While the basic calorimetric performance is comparable to the state of the art, the technology will clearly be superior in terms of two-shower separation. Much more detailed analysis of e.g. the event-by-event behaviour of shower shapes can be performed and will be the scope of future publications.

\section*{Acknowledgments}
We would like to thank J.A.~Hasenbichler for providing the results of the TCAD simulations, J.~Schambach for the multi-channel transition board design, and M.~Bonora and M.~Lupi for help in installation/implementation and adaptation of FPGA firmware and software. We also thank the test beam coordinators and the supporting people at DESY and the CERN SPS for the usage of the test beam, and M.~Stanitzki for useful discussions. This work was partially supported by the Nederlandse Organisatie voor Wetenschappelijk Onderzoek (NWO), Netherlands.

\section*{References}

\bibliography{pisameet2021-peitzmann}

\end{document}